\documentclass[prd,nofootinbib,showpacs,superscriptaddress, preprint]{revtex4-1}
\usepackage[T1]{fontenc}
\usepackage{amsmath,amssymb}
\usepackage{epsfig}
\usepackage{graphicx}
\usepackage[usenames,dvipsnames]{color}
\usepackage{slashed}
\usepackage[colorlinks,citecolor=blue]{hyperref}
\usepackage{comment}
\usepackage{bbold}
\usepackage{color}

\usepackage{soul,xcolor}
\usepackage{tikz-feynman}
\tikzfeynmanset{compat=1.1.0}

\usepackage{tikzsymbols}
\begin{document}
\title{Two-component Dark Matter with co-genesis of Baryon Asymmetry of the Universe}

\author{Debasish Borah}%
 \email{dborah@iitg.ac.in}
\affiliation{%
 Department of Physics, Indian Institute of Technology, Guwahati, Assam 781039, India
 }%
\author{Arnab Dasgupta}%
 \email{arnabdasgupta@protonmail.ch}
\affiliation{%
 School of Liberal Arts, Seoul-Tech, Seoul 139-743, Korea}%
\author{Sin Kyu Kang}%
 \email{skkang@seoultech.ac.kr}
\affiliation{%
 School of Liberal Arts, Seoul-Tech, Seoul 139-743, Korea}%
 
 \begin{abstract}
We discuss the possibility of realising a two-component dark matter (DM) scenario where the two DM candidates differ from each other by virtue of their production mechanism in the early universe. One of the DM candidates is thermally generated in a way similar to the weakly interacting massive particle (WIMP) paradigm where the DM abundance is governed by its freeze-out while the other candidate is produced only from non-thermal contributions similar to freeze-in mechanism. We discuss this in a minimal extension of the standard model where light neutrino masses arise radiatively in a way similar to the scotogenic models with DM particles going inside the loop. The lepton asymmetry is generated at the same time from WIMP DM annihilations as well as partially from the mother particle for non-thermal DM.  
This can be achieved while satisfying the relevant experimental bounds, and keeping the scale of leptogenesis or the thermal DM mass  as low as 3 TeV, well within present experimental reach. In contrast to the TeV scale thermal DM mass, the non-thermal DM can be as low as a few keV, giving rise to the possibility of a sub-dominant warm dark matter (WDM) component that can have interesting consequences on structure formation. The model also has tantalizing prospects of being detected at ongoing direct detection experiments as well as the ones looking for charged lepton flavour violating process like $\mu \rightarrow e \gamma$.

\end{abstract}
\maketitle

\section{\label{sec:level1}Introduction}
There have been irrefutable amount of evidences suggesting the presence of a mysterious, non-luminous, collisionless and non-baryonic form of matter in the present universe \cite{Tanabashi:2018oca}. The hypothesis for existence of this form of matter, more popularly known as dark matter (DM) due to its non-luminous nature, is strongly backed by early galaxy cluster observations~\cite{Zwicky:1933gu}, observations of galaxy rotation curves~\cite{Rubin:1970zza}, the more recent observation of the bullet cluster~\cite{Clowe:2006eq} and the latest cosmological data provided by the Planck satellite~\cite{Aghanim:2018eyx}. The latest data from the Planck satellite suggest that around $27\%$ of the present universe's energy density is in the form of dark matter. In terms of density parameter $\Omega$ and $h = \text{(Hubble Parameter)}/(100 \;\text{km} \text{s}^{-1} \text{Mpc}^{-1})$, the present dark matter abundance is conventionally reported as
\begin{equation}
\Omega_{\text{DM}} h^2 = 0.120\pm 0.001
\label{dm_relic}
\end{equation}
at 68\% CL~\cite{Aghanim:2018eyx}. While such astrophysics and cosmology based experiments are providing such evidence suggesting the presence of dark matter at regular intervals in the last several decades, there is hardly anything known about the particle nature of it. The requirements which a particle dark matter candidate has to satisfy, as pointed out in details by the authors of \cite{Taoso:2007qk} rule out all the standard model (SM) particles from being DM candidates. While the neutrinos in the SM come very close to satisfying these requirements, they have tiny abundance in the present universe. Apart from that, they have a large free streaming length (FSL) due to their relativistic nature and give rise to hot dark matter (HDM), ruled out by observations. This has led to a plethora of beyond standard model (BSM) scenarios proposed by the particle physics community to account for dark matter in the universe. Most of these BSM scenarios are based on a popular formalism known as the weakly interacting massive particle (WIMP) paradigm. In this formalism, a particle dark matter candidate having mass around the electroweak scale and having electroweak type couplings to SM particles can give rise to the correct relic abundance in the present epoch, a remarkable coincidence often referred to as the \textit{WIMP Miracle}~\cite{Kolb:1990vq}. Since the mass is around the electroweak corner and couplings to the SM particles are sizeable, such DM candidates are produced thermally in the early universe followed by its departure from chemical equilibrium leading to its freeze-out. Such DM candidates typically become non-relativistic shortly before the epoch of freeze-out and much before the epoch of matter-radiation equality. Such DM candidates are also categorised as cold dark matter (CDM).

The CDM candidates in the WIMP paradigm have very good direct detection prospects due to its sizeable interaction strength with SM particles and hence can be observed at ongoing and future direct search experiments \cite{panda17, Tan:2016zwf, Aprile:2017iyp, Akerib:2016vxi, Akerib:2015cja, Aprile:2015uzo, Aalbers:2016jon, Liu:2017drf}. However, no such detection has yet been done casting doubts over the viability of such DM paradigms. This has also motivated the particle physics community to look for other alternatives to WIMP paradigm. Although such null results could indicate a very constrained region of WIMP parameter space, they have also motivated the particle physics community to look for beyond the thermal WIMP paradigm where the interaction scale of DM particle can be much lower
than the scale of weak interaction i.e.\,\,DM may be more feebly
interacting than the thermal WIMP paradigm. This falls under the ballpark of non-thermal DM \cite{Hall:2009bx}. In this scenario, the initial number density of DM in the early Universe is
negligible and it is assumed that the interaction strength of DM
with other particles in the thermal bath is so feeble that
it never reaches thermal equilibrium at any epoch in the early Universe. In this set up,
DM is mainly produced from the out of equilibrium decays
of some heavy particles in the plasma. It can also be produced
from the scatterings of bath particles, however if same couplings
are involved in both decay as well as scattering processes
then the former has the dominant contribution to DM relic density
over the latter one \cite{Hall:2009bx}.
The production mechanism for non-thermal DM
is known as freeze-in and the candidates of non-thermal DM produced
via freeze-in are often classified into a group called
Freeze-in (Feebly interacting) massive particle (FIMP). For a recent review of this DM paradigm, please see \cite{Bernal:2017kxu}. Interestingly, such non-thermal DM candidates can have a wide range of allowed masses, well beyond the typical WIMP regime. The possibility of a light DM candidate have interesting implications for astrophysical structure formation in the universe. Although a light DM candidate like SM neutrinos which constitute HDM is already ruled out and CDM is one of the most well studied scenario (specially within the context of WIMP paradigm), there also exists an intermediate possibility where DM remains mildly relativistic at the epoch of matter-radiation equality. Consequently, the free streaming length of such candidates fall in between the large FSL of HDM and small FSL of CDM. Such DM candidates which can be kept at intermediate stage between HDM and CDM are typically referred to as warm dark matter (WDM). WDM candidates have typical masses in keV range, in contrast to typical mass of HDM in sub-eV mass and CDM with GeV-TeV scale masses. For a recent review on keV scale singlet fermion as WDM, please have a look at \cite{Adhikari:2016bei}. Although such WDM candidates may not be as motivating as WIMP or typical CDM candidates from direct search point of view, there are strong motivations from astrophysics point of view. Typical WDM scenarios can provide a solution to several small scale structure problems faced by CDM paradigm. The missing satellite problem, too big to fail problem fall in the list of such small structure problems, a recent review of which can be found in \cite{Bullock:2017xww}. The above mentioned classification of HDM, CDM and WDM is primarily based on their FSL, typically equal to the distance for which the DM particles can propagate freely without interacting. Typically, the free streaming length $\lambda_{\text{FS}} =0.1$ Mpc, about the size of a dwarf galaxy, acts as a boundary line between HDM ($\lambda_{\text{FS}} >0.1$ Mpc) and WDM ($\lambda_{\text{FS}} <0.1$ Mpc). For CDM, on the other hand, the FSL are considerably smaller than this value. Therefore, CDM structures keep forming till scales as small as the solar system which gives rise to disagreement with observations at small scales \cite{Bullock:2017xww}. HDM, on the other hand, erases all small scale structure due to its large free streaming length, disfavouring the bottom up approach of structure formation. WDM can therefore act as a balance between the already ruled out HDM possibility and the CDM paradigm having issues with small scale structures. More details about the calculation of FSL can be found in \cite{Boyarsky:2008xj, Merle:2013wta}. We show that our non-thermal DM candidate can be a keV scale fermion which can give rise to a sub-dominant WDM component. Such mixed CDM and WDM type hybrid DM scenario was also considered in some recent works \cite{Borah:2017hgt, DuttaBanik:2016jzv}. However, our model is not restrictive to such combinations as we show that the non-thermal DM candidate can have masses in the keV-GeV range as well.

Apart from the mysterious $27\%$ of the universe in the form of unknown DM, the visible sector making up to $5\%$ of the universe also creates a puzzle. This is due to the asymmetric nature of the visible sector. The visible or baryonic part of the universe has an abundance of baryons over anti-baryons. This is also quoted as baryon to photon ratio $(n_{B}-n_{\bar{B}})/n_{\gamma} \approx 10^{-10}$ which is rather large keeping in view of the large number density of photons. If the universe is assumed to start in a symmetric manner at the big bang epoch which is a generic assumption, there has to be a dynamical mechanism that can lead to a baryon asymmetric universe at present epoch. The requirements such a dynamical mechanism needs to satisfy were put forward by Sakharov more than fifty years ago, known as the Sakharov's conditions~\cite{Sakharov:1967dj}: baryon number (B) violation, C and CP violation and departure from thermal equilibrium. Unfortunately, all these requirements can not be fulfilled in the required amount within the framework of the SM, again leading to several BSM scenarios. out of equilibrium decay of a heavy particle leading to the generation of baryon asymmetry has been a very well known mechanism for baryogenesis \cite{Weinberg:1979bt, Kolb:1979qa}. One interesting way to implement such a mechanism is leptogenesis \cite{Fukugita:1986hr} where a net leptonic asymmetry is generated first which gets converted into baryon asymmetry through $B+L$ violating EW sphaleron transitions. The interesting feature of this scenario is that
the required lepton asymmetry can be generated within the framework of the seesaw mechanism 
\cite{Minkowski:1977sc, Mohapatra:1979ia, Yanagida:1979as, GellMann:1980vs, Glashow:1979nm, Schechter:1980gr} that explains the origin of tiny neutrino masses \cite{Tanabashi:2018oca}, another observed phenomena which the SM fails to address.

Although the explanation for dark matter, baryon asymmetry of the universe and origin of neutrino mass can arise independently in different BSM frameworks, it is interesting, economical and predictive to consider a common framework for their origin. In fact a connection between DM and baryons appears to be a natural possibility  to understand their same order of magnitude abundance $\Omega_{\rm DM} \approx 5 \Omega_{B}$. Discarding the possibility of any numerical coincidence, one is left with the task of constructing theories that can relate the origin of these two observed phenomena in a unified manner. There have been several proposals already which mainly fall into two broad categories. In the first one, the usual mechanism for baryogenesis  is extended to apply to the dark sector which is also asymmetric \cite{Nussinov:1985xr, Davoudiasl:2012uw, Petraki:2013wwa, Zurek:2013wia}. The second one is to produce such asymmetries through annihilations \cite{Yoshimura:1978ex, Barr:1979wb, Baldes:2014gca} where one or more particles involved in the annihilations eventually go out of thermal equilibrium in order to generate a net asymmetry. 
The so-called WIMPy baryogenesis \cite{Cui:2011ab, Bernal:2012gv, Bernal:2013bga} belongs to this category, where a dark matter particle freezes out to generate its own relic abundance and then
an asymmetry in the baryon sector is produced from DM annihilations. The idea extended to leptogenesis is called WIMPy leptogenesis \cite{Kumar:2013uca, Racker:2014uga, Dasgupta:2016odo, Borah:2018uci}. Motivated by all these, we propose a scenario where the DM sector is a hybrid of one thermal and one non-thermal components while the thermal DM annihilations play a dominant role in creating a leptonic asymmetry which gets converted into baryon asymmetry eventually, after electroweak phase transition. The non-thermal DM can also be at keV scale giving rise to the possibility of WDM which can have interesting consequences at astrophysical structure formation as well as DM indirect detection experiments. The neutrino mass arises at one loop level where the dark sector particles take part in the loop mediation.

This paper is organised as follows. In section \ref{sec1} we discuss our model followed by the origin of neutrino mass in section \ref{sec2}. In section \ref{sec3} we describe the co-genesis of WIMP, FIMP and lepton asymmetry followed by relevant constraints from direct detection and lepton flavour violation in section \ref{sec4}. We then discuss our results in section \ref{sec5} and finally conclude in section \ref{sec6}.

\section{The Model}
\label{sec1}
We consider a minimal extension of the SM by two different types of singlet fermions and three different types of scalar fields shown in table \ref{tab1a}, \ref{tab2a} respectively. To achieve the desired interactions of these new fields among themselves as well as with the SM particles, we consider additional discrete symmetries $\mathbb{Z}_2 \times \mathbb{Z}^{\prime}_2$. While one such $\mathbb{Z}_2$ symmetry is enough to accommodate DM, radiative neutrino mass as well as generation of lepton asymmetry from DM annihilation in a way similar to what we achieve in a version of scotogenic model \cite{Borah:2018uci}, the other discrete symmetry $\mathbb{Z}^{\prime}_2$ is required in order to have the desired couplings of FIMP DM. To prevent tree level interaction between FIMP DM and SM leptons through $\bar{L} \tilde{H} \chi$ (needed to avoid the decay of $\chi$ to light SM particles), we have introduced this another discrete symmetry $\mathbb{Z}^{\prime}_2$ under which $\chi, \phi$ and another singlet scalar $\phi^{\prime}$ are odd. If $\phi^{\prime}$ acquires a non-zero vacuum expectation value (vev), it can lead to one loop mixing between neutrinos and non-thermal DM. This possibility is shown in table \ref{tab1a}, \ref{tab2a}.

\begin{table}
\begin{center}
\begin{tabular}{|c|c|}
\hline
Particles & $SU(3)_c \times SU(2)_L \times U(1)_{Y} \times \mathbb{Z}_2 \times \mathbb{Z}^{\prime}_2$   \\
\hline
$Q_L$ & $(3,2, \frac{1}{3}, +, +)$ \\
$u_R$ & $(3^*,1,\frac{4}{3}, +, +)$  \\
$d_R$ & $(3^*,1,-\frac{2}{3}, +, +)$  \\
$\ell_L$ & $(1,2,-1, +, +)$  \\
$\ell_R$ & $(1,1,-2, +, +)$ \\
$\chi$ & $(1,1,0, +, -)$ \\
$N$ & $(1,1,0, -, +)$ \\
\hline
\end{tabular}
\end{center}
\caption{Fermion content of the model}
\label{tab1a}
\end{table}

\begin{table}
\begin{center}
\begin{tabular}{|c|c|}
\hline
Particles & $SU(3)_c \times SU(2)_L \times U(1)_{Y} \times \mathbb{Z}_2 \times \mathbb{Z}^{\prime}_2$   \\
\hline
$H$ & $(1,2, 1, +, +)$ \\
$\eta$ & $(1,2, 1, -, +)$ \\
$\phi$ & $(1,1,0, -, -)$ \\
$\phi^{\prime}$ & $(1,1,0, +, -)$ \\
\hline
\end{tabular}
\end{center}
\caption{Scalar content of the model}
\label{tab2a}
\end{table}

The relevant part of the Yukawa Lagrangian is
\begin{align}
{\cal L} & \supset  \frac{1}{2}(M_N)_{ij} N_iN_j + \frac{1}{2} m_{\chi} \chi \chi+y_{ij} \, \bar{L}_i \tilde{\eta} N_j  +y^{\prime}_i \phi \chi N_i+ \text{h.c.}.
\label{yukawa2} 
\end{align}
The scalar potential is
\begin{equation}
V = V_{H \eta} + V_{H \phi} + V_{\eta \phi}
\end{equation}
where
\begin{align}
V_{H \eta} &=  \mu_H^2|H|^2 +\mu_{\eta}^2|\eta|^2+\frac{\lambda_1}{2}|H|^4+\frac{\lambda_2}{2}|\eta|^4+\lambda_3|H|^2|\eta|^2 \nonumber \\
& +\lambda_4|H^\dag \eta|^2 + \{\frac{\lambda_5}{2}(H^\dag \eta)^2 + \text{h.c.}\},
\label{scalar1a}
\end{align}
\begin{align}
V_{H \phi} & =  \mu_{\phi}^2\phi^2+\frac{\lambda_6}{2} \phi^4+\lambda_7|H|^2 \phi^2 + \mu_{\phi^{\prime}}^2 (\phi^{\prime})^2+\frac{\lambda^{\prime}_6}{2} (\phi^{\prime})^4+\lambda^{\prime}_7|H|^2 (\phi^{\prime})^2
\label {scalar2a}
\end{align}
\begin{align}
V_{\eta \phi} & =  \lambda_8 |\eta|^2 \phi^2 +  \lambda_9 \eta^{\dagger} H \phi \phi^{\prime} +  \lambda^{\prime}_8 |\eta|^2 (\phi^{\prime})^2 +  \lambda^{\prime}_9 \phi^2 (\phi^{\prime})^2
\label {scalar3a}
\end{align}

Since we require the SM Higgs and the singlet scalar $\phi^{\prime}$ to acquire non-zero vev as 
$$ \langle H \rangle=\left( 0,  \;\;  \frac{ v}{\sqrt 2} \right)^T, \;\; \langle \phi^{\prime} \rangle = \frac{u}{\sqrt 2} $$
we minimise the above scalar potential with respect to these two fields and find the following minimisation conditions.
\begin{align}
-\mu^2_H = \frac{\lambda_1}{2} v^2 + \frac{\lambda^{\prime}_7}{2} u^2 \nonumber \\
-\mu^2_{\phi^{\prime}} = \frac{\lambda^{\prime}_6}{2} u^2 + \frac{\lambda^{\prime}_7}{2} v^2
\end{align}
The corresponding mass squared matrix is 
\begin{equation}
M^2_{H \phi^{\prime}} =
\left(\begin{array}{cc}
\ \lambda_1 v^2 & \lambda^{\prime}_7 \frac{vu}{2} \\
\ \lambda^{\prime}_7 \frac{vu}{2} & \lambda^{\prime}_6 u^2
\end{array}\right)
\end{equation}
This will give rise to a mixing between the SM like Higgs and a singlet scalar given by
\begin{equation}
\tan{2\theta_1} \approx 2\sin{\theta_1} \approx 2\theta_1 = \frac{\lambda^{\prime}_7 vu}{\lambda^{\prime}_6 u^2-\lambda_1 v^2} \approx \frac{\lambda^{\prime}_7 v}{\lambda^{\prime}_6 u}
\end{equation}
where in the last step we have assumed a hierarchy $u \gg v$. The mass eigenstates corresponding to the charged and pseudo-scalar components of $\eta$ are
\begin{eqnarray}
m_{\eta^\pm}^2 &=& \mu_{\eta}^2 + \frac{1}{2}\lambda_3 v^2 , \nonumber\\
m_{\eta_I}^2 &=& \mu_{\eta}^2 + \frac{1}{2}(\lambda_3+\lambda_4-\lambda_5)v^2=m^2_{\eta^\pm}+
\frac{1}{2}\left(\lambda_4-\lambda_5\right)v^2.
\label{mass_relation}
\end{eqnarray}
The neutral scalar component of $\eta$ and $\phi$ mix with each other resulting in the following mass squared matrix.
\begin{equation}
M^2_{\eta \phi} =
\left(\begin{array}{cc}
\  \mu_{\eta}^2 + \frac{1}{2}(\lambda_3+\lambda_4+\lambda_5)v^2 +\frac{\lambda^{\prime}_8}{2}u^2 & \lambda_9 \frac{vu}{4} \\
\ \lambda_9 \frac{vu}{4}  & \mu^2_{\phi}+\frac{\lambda_7}{2}v^2 + \frac{\lambda^{\prime}_9}{2}u^2
\end{array}\right),
\end{equation}
which can be diagonalized by a $2\times 2 $ unitary matrix with the mixing angle given by
\begin{equation}
\tan{2\theta_2} \approx 2 \sin{\theta_2} \approx 2 \theta_2 = \frac{\lambda_9 vu}{2( \mu^2_{\phi}+\frac{\lambda_7}{2}v^2 + \frac{\lambda^{\prime}_9}{2}u^2-\mu_{\eta}^2 - \frac{1}{2}(\lambda_3+\lambda_4+\lambda_5)v^2 -\frac{\lambda^{\prime}_8}{2}u^2)}
\end{equation}

\begin{figure}
\centering
\begin{tabular}{lcr}
\begin{tikzpicture}[/tikzfeynman/small]
\begin{feynman}
\vertex (i){$N$};
\vertex [right = 1.13cm of i] (v);
\vertex [above = 0.8cm of v] (v1);
\vertex [right = 0.8cm of v1](j){$\phi$};
\vertex [below = 1.6cm of j](k){$\chi$};
\diagram*[small]{(i)--[fermion](v),(v)--[scalar](j),(v)--[fermion](k)};
\end{feynman}
\end{tikzpicture}
&
\begin{tikzpicture}[/tikzfeynman/small]
\begin{feynman}
\vertex (i){$\nu,l^{\pm}$};
\vertex [below = 1.6cm of i] (j){$\eta^0,\eta^\mp$};
\vertex [right = 0.8cm of i] (v1);
\vertex [below = 0.8cm of v1] (va);
\vertex [right = 0.8cm of va] (vb);
\vertex [above = 0.8cm of vb] (v2);
\vertex [right = 0.5cm of v2] (k){$\phi$};
\vertex [below = 1.6cm of k] (l){$\chi$};
\diagram*[small]{(i)--[fermion](va),(j)--[anti charged scalar](va),(va)--[fermion,edge label=$N$](vb),(vb)--[fermion](l),(vb)--[scalar](k)};
\end{feynman}
\end{tikzpicture}
&
\begin{tikzpicture}[/tikzfeynman/small]
\begin{feynman}
\vertex (i){$\eta^0,\eta^\mp$};
\vertex [right = 1.5cm of i] (v1);
\vertex [above right = 0.8cm of v1] (a){$\nu,l^\mp$};
\vertex [right = 1.2cm of v1] (v2);
\vertex [above right = 0.8cm of v2] (b){$\phi$};
\vertex [right = 0.8cm of v2] (c){$\chi$};
\diagram*[small]{(i)--[charged scalar](v1),(v1)--[anti fermion](a),(v1)--[fermion,edge label'=$N_i$](v2),(v2)--[fermion](c),(v2)--[scalar](b)};
\end{feynman}
\end{tikzpicture}
\end{tabular}
\caption{Feynman diagrams corresponding to the production of non-thermal DM $\chi$.}
\label{fig:fimp}
\end{figure}
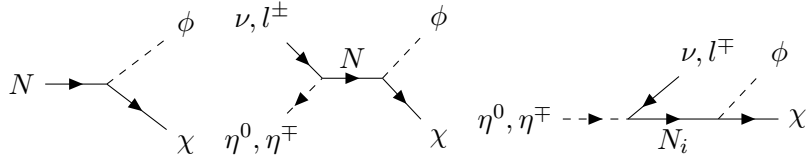

Considering $ \lvert m_{\eta} - m_{\phi} \rvert < m_{\chi}$, we can prevent the three body decays $\eta \rightarrow \phi \chi \nu$ or $\phi \rightarrow \eta \chi \nu$. Even if we allow such three body decays, they will be phase space suppressed compared to two body decays which contribute to the production of non-thermal DM $\chi$ which we will discuss shortly. We also consider the mixing between $\eta-\phi$ to be non-zero so that the thermal DM is an admixture of singlet and doublet scalars\footnote{We will use the notation DM to denote them in our analysis.}. This has crucial implications for DM phenomenology as well as leptogenesis as we discuss below. Assuming $m_{\chi} \sim$ keV-GeV, we consider its production mechanisms which also have the potential to produce a lepton asymmetry. The relevant diagrams for producing lepton asymmetry and non-thermal DM are shown in  FIG. \ref{fig:coasym} and \ref{fig:fimp}, respectively.

\begin{figure}
\centering
\begin{tabular}{lr}
\begin{tikzpicture}[/tikzfeynman/small]
\begin{feynman}
\vertex (i){$\eta^-$};
\vertex [below = 2.cm of i] (j){$N_i$};
\vertex [below right= 1.414cm of i] (v1);
\vertex [right = 1.cm of v1] (v2);
\vertex [right = 3.cm of i] (m){$X$};
\vertex [below = 2.cm of m] (o){$L_\alpha$};
\diagram*[small]{(i) -- [charged scalar] (v1),(v1) -- [fermion,edge label = $l_\alpha$] (v2),(v2) -- [ scalar] (m),(v2) -- [fermion] (o),(j) -- [anti fermion] (v1)};
\end{feynman}
\end{tikzpicture}
& 
\begin{tikzpicture}[/tikzfeynman/small]
\begin{feynman}
\vertex (i){$\eta^-$};
\vertex [right = 3.cm of i] (j){$X$};
\vertex [below = 1.5cm of i] (k){$N_i$};
\vertex [below = 1.5cm of j] (l){$L_\alpha$};
\vertex [right = 1.5cm of i] (v1);
\vertex [below = 1.5cm of v1] (v2);
\diagram*[small]{(i) -- [charged scalar] (v1),(v1) -- [charged scalar,edge label = $\eta$] (v2),(v1) -- [scalar] (j),(v2) -- [ fermion] (l),(k) -- [anti fermion] (v2)};
\end{feynman}
\end{tikzpicture}
\\ 
\begin{tikzpicture}[/tikzfeynman/small]
\begin{feynman}
\vertex (i){$\eta^-$};
\vertex [right = 4.cm of i] (j){$X$};
\vertex [below = 2.cm of i] (k){$N_i$};
\vertex [below = 2cm of j] (l){$L_\alpha$};
\vertex [right = 1.cm of k] (k1);
\vertex [left = 1.cm of l] (l1);
\vertex [right = 2.cm of i] (v1);
\vertex [below = 2.cm of v1] (v2);
\vertex [below = 1.cm of v1] (v3);
\vertex [left = 0.5cm of v3] (v4);
\vertex [below = 1.5cm of v4] (v5);
\diagram*[small]{(i) -- [charged scalar] (v1),(v1) -- [charged scalar,edge label = $\eta$] (v3),(k) -- [anti fermion] (k1),(k1) -- [anti fermion,edge label=$l_\beta$] (v3),(k1) -- [charged scalar,edge label'=$\eta$] (l1),(v3) -- [ majorana,edge label=$N_j$] (l1),(v1) -- [scalar] (j),(l1) -- [fermion] (l),(v4) -- [red,scalar](v5)};
\end{feynman}
\end{tikzpicture}
&
\begin{tikzpicture}[/tikzfeynman/small]
\begin{feynman}
\vertex (i){$\eta^-$};
\vertex [below = 1.6cm of i] (j){$N_i$};
\vertex [right = 1.4cm of i] (i1);
\vertex [below = 1.6cm of i1] (j1);
\vertex [right = 1.cm of i] (l1);
\vertex [below = 1.6cm of l1] (l3);
\vertex [right = 1.cm of l1] (l2);
\vertex [below = 0.8cm of l2] (v1);
\vertex [right = 4.2cm of i] (k){$X$};
\vertex [right = 1.2cm of v1] (v2);
\vertex [below = 1.6cm of k] (m){$L_\alpha$};
\diagram*[small]{(i) -- [charged scalar] (l1),(l1) -- [majorana,edge label=$N_j$](v1),(v1)--[scalar](l3),(l3)--[anti fermion,edge label = $l_\beta$](l1),
(j)--[fermion](l3),(v1) -- [fermion,edge label = $l_\alpha$](v2),(v2)--[scalar](k),(v2)--[fermion](m),(i1)--[red,scalar](j1)};
\end{feynman}
\end{tikzpicture}
\end{tabular}
\caption{Feynman diagrams contributing to $\langle\sigma v\rangle_{\eta N_i \rightarrow X L}$ and the interference term $\epsilon$. Here $X \equiv h,\gamma,W^{\pm},Z$.}
\label{fig:coasym}
\end{figure}
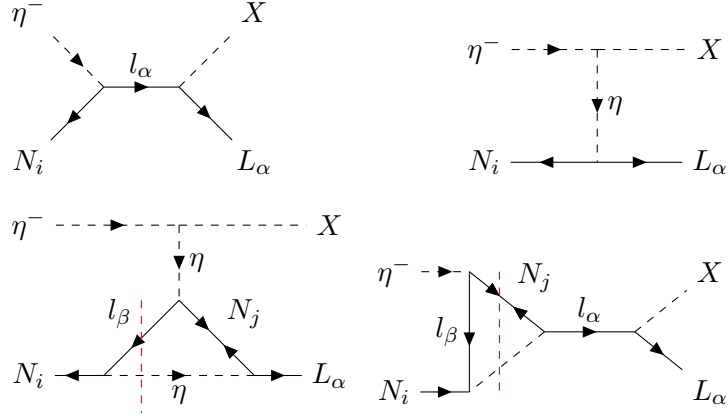

we implement the model in \texttt{SARAH 4} \cite{Staub:2013tta} and extract the thermally averaged annihilation rates from \texttt{micrOMEGAs 4.3} \cite{Barducci:2016pcb} to use while solving the relevant Boltzmann equations to be discussed below.
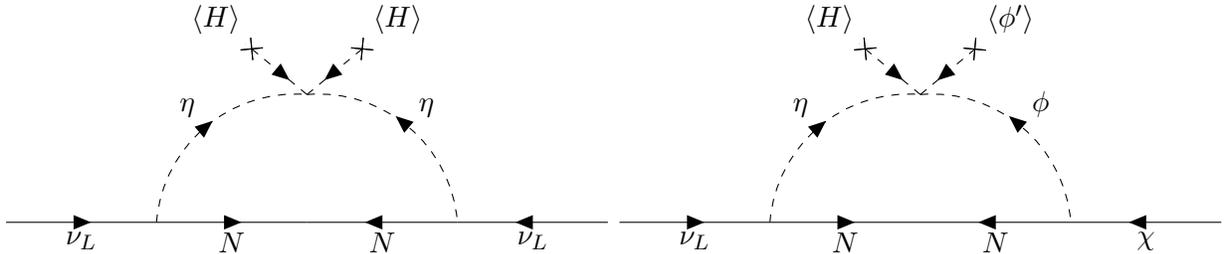
\begin{figure}
\centering
\begin{tabular}{lr}
\begin{tikzpicture}
\begin{feynman}
\vertex (i1);
\vertex [right=2cm of i1] (a);
\vertex [right=2cm of a] (b);
\vertex [right=2cm of b] (c);
\vertex [right=2cm of c] (d);
\vertex [above=1.7cm of b] (tb1);
\vertex [above=1.cm of tb1] (z2);
\vertex [left=0.75cm of z2] (wh) {$\left\langle H\right\rangle$};
\vertex [right=0.75cm of z2] (wp) {$\langle H \rangle$};
\diagram* {
i1 -- [fermion, edge label'=$\nu_{L}$] (a) -- [fermion, edge label'=$N$] (b),(b) -- [anti fermion, edge label'=$N$] (c) -- [anti fermion, edge label'=$\nu_{L}$] (d),
a -- [charged scalar, quarter left, edge label=$\eta$] (tb1),
tb1 -- [anti charged scalar, insertion=0.9] (wh),
tb1 -- [anti charged scalar, insertion=0.9] (wp),
tb1 -- [anti charged scalar, quarter left, edge label=$\eta$] (c),
};
\end{feynman}
\end{tikzpicture}
&
\begin{tikzpicture}
\begin{feynman}
\vertex (i1);
\vertex [right=2cm of i1] (a);
\vertex [right=2cm of a] (b);
\vertex [right=2cm of b] (c);
\vertex [right=2cm of c] (d);
\vertex [above=1.7cm of b] (tb1);
\vertex [above=1.cm of tb1] (z2);
\vertex [left=0.75cm of z2] (wh) {$\left\langle H\right\rangle$};
\vertex [right=0.75cm of z2] (wp) {$\langle \phi^\prime \rangle$};
\diagram* {
i1 -- [fermion, edge label'=$\nu_{L}$] (a) -- [fermion, edge label'=$N$] (b),b -- [anti fermion, edge label'=$N$] (c) -- [anti fermion, edge label'=$\chi$] (d),
a -- [charged scalar, quarter left, edge label=$\eta$] (tb1),
tb1 -- [anti charged scalar, insertion=0.9] (wh),
tb1 -- [anti charged scalar, insertion=0.9] (wp),
tb1 -- [anti charged scalar, quarter left, edge label=$\phi$] (c),
};
\end{feynman}
\end{tikzpicture}
\end{tabular}
\caption{Radiative light neutrino mass and mixing of non-thermal DM with light neutrinos}
\label{numassmixing}
\end{figure}

\section{Neutrino Mass}
\label{sec2}
As can be noticed from the particle content of the model and the Yukawa Lagrangian mentioned above, light neutrino mass does not arise at tree level as long as one of the discrete symmetries namely, $\mathbb{Z}_2$ remains unbroken. At one loop level however, one can have light neutrino masses originating from the diagram shown in the left panel of FIG. \ref{numassmixing}. This is same as the way light neutrino masses are generated in scotogenic model proposed by Ma \cite{Ma:2006km}. The one-loop expression for neutrino mass is
\begin{equation}
(m_{\nu})_{ij}=\sum_{k} \frac{y_{ik} y_{kj} M_{k}}{32 \pi^{2}} \left[\frac{m_{\eta R}^{2}}{m_{\eta R}^{2}-M_{k}^{2}} \log{\left(\frac{m_{\eta R}^{2}}{M_{k}^{2}}\right)}-\frac{m_{\eta I}^{2}}{m_{\eta I}^{2}-M_{k}^{2}} \log\left(\frac{m_{\eta I}^{2}}{M_{k}^{2}}\right)\right]
\label{n_mass}
\end{equation}
where $M_{k}$ is the right handed neutrino mass. The above Eq.~\eqref{n_mass} equivalently can be written as
\begin{equation}
(m_{\nu})_{ij} \equiv (y^{T}\Lambda y)_{ij}
\end{equation}
where $\Lambda$ can be defined as,
\begin{equation}
\Lambda_{k} = \frac{M_{k}}{32 \pi^{2}} \left[\frac{m_{\eta R}^{2}}{m_{\eta R}^{2}-M_{k}^{2}} \log{\left(\frac{m_{\eta R}^{2}}{M_{k}^{2}}\right)}-\frac{m_{\eta I}^{2}}{m_{\eta I}^{2}-M_{k}^{2}} \log\left(\frac{m_{\eta I}^{2}}{M_{k}^{2}}\right)\right].
\end{equation}

In order to incorporate the constraints from neutrino oscillation data on three mixing angles and two mass squared differences, it is often useful to express these Yukawa couplings in terms of light neutrino parameters. This is possible through the Casas-Ibarra (CI) parametrisation \cite{Casas:2001sr} extended to radiative seesaw model \cite{Toma:2013zsa} which allows us to write the Yukawa couplings as
\begin{equation}
y=\sqrt{\Lambda}^{-1} R \sqrt{m^{\rm diag}_{\nu}} U_{\rm PMNS}^{\dagger}.
\end{equation}
Here
$m^{\rm diag}_{\nu} = {\rm diag}(m_1, m_2, m_3)$ is the diagonal light 
neutrino mass matrix and $R$ can be a complex orthogonal matrix in general with $RR^{T}=\mathbb{1}$ which we have taken it to be a general,  this $3\times3$ orthogonal matrix $R$ can be parametrised by three complex parameters of type $\theta_{\alpha \beta} = \theta^R_{\alpha \beta} + i\theta^I_{\alpha \beta}, \theta^R_{\alpha \beta} \in [0, 2\pi], \theta^I_{\alpha \beta} \in \mathbb{R}$ \cite{Ibarra:2003up} \footnote{For some more discussions on different possible structure of this matrix and implications on a particular leptogenesis scenario in this model, we refer to the recent work \cite{Mahanta:2019gfe}.}.
In general, the orthogonal matrix $R$ for $n$ flavours can be product of $^nC_2$ number of rotation matrices of type
\begin{align}
    R_{\alpha \beta} &= \begin{pmatrix} \cos{(\theta^R_{\alpha \beta} + i\theta^I_{\alpha \beta})} & \cdots & \sin{(\theta^R_{\alpha \beta} + i\theta^I_{\alpha \beta})} \\
    \vdots & \ddots & \vdots \\
  - \sin{(\theta^R_{\alpha \beta} + i\theta^I_{\alpha \beta})} & \cdots & \cos{(\theta^R_{\alpha \beta} + i\theta^I_{\alpha \beta})} \end{pmatrix},
\end{align}
with rotation in the $\alpha-\beta$ plane and dots stand for zero. For example, taking $\alpha=1, \beta=2$ we have
\begin{align}
  R_{12} &= \begin{pmatrix} \cos{(\theta^R_{12} + i\theta^I_{12})} &  \sin{(\theta^R_{12} + i\theta^I_{12})} & 0 \\
  -\sin{(\theta^R_{12} + i\theta^I_{12})} &  \cos{(\theta^R_{12} + i\theta^I_{12})} & 0 \\
  0 &  0 & 1\end{pmatrix}.
\end{align} 
We see that CP phases in $U$ do not contribute to $\epsilon_{N_i\eta}$ given in eq.\eqref{eq:asymB}, but
complex variables in the orthogonal matrix $R$ can lead to non-vanishing 
value of $\epsilon_{N_i\eta}$. This is similar to leptogenesis from pure 
decay in this model \cite{Hugle:2018qbw} where, in the absence of flavour 
effects, the orthogonal matrix $R$ played a crucial role. The matrix denoted by $U_{\rm PMNS}$ is the 
Pontecorvo-Maki-Nakagawa-Sakata (PMNS) leptonic mixing matrix
\begin{equation}
U_{\text{PMNS}} = U^{\dagger}_l U_L.
\label{pmns0}
\end{equation}
If the charged lepton mass matrix is diagonal or equivalently, $U_L = \mathbb{1}$, then the PMNS mixing matrix is identical to the diagonalising matrix of neutrino mass matrix. 
The PMNS mixing matrix can be parametrised as
\begin{equation}
U_{\text{PMNS}}=\left(\begin{array}{ccc}
c_{12}c_{13}& s_{12}c_{13}& s_{13}e^{-i\delta}\\
-s_{12}c_{23}-c_{12}s_{23}s_{13}e^{i\delta}& c_{12}c_{23}-s_{12}s_{23}s_{13}e^{i\delta} & s_{23}c_{13} \\
s_{12}s_{23}-c_{12}c_{23}s_{13}e^{i\delta} & -c_{12}s_{23}-s_{12}c_{23}s_{13}e^{i\delta}& c_{23}c_{13}
\end{array}\right) U_{\text{Maj}}
\label{matrixPMNS}
\end{equation}
where $c_{ij} = \cos{\theta_{ij}}, \; s_{ij} = \sin{\theta_{ij}}$ and $\delta$ is the leptonic Dirac CP phase. 
The diagonal matrix $U_{\text{Maj}}=\text{diag}(1, e^{i\alpha}, e^{i\beta})$  contains the Majorana 
CP phases $\alpha, \beta$ which remain undetermined at neutrino oscillation experiments. We summarise the $3\sigma$ global fit values in table \ref{tabglobalfit} from the recent analysis \cite{Esteban:2018azc}, which we use in our subsequent analysis.

The diagram on right panel of FIG. \ref{numassmixing} gives rise to radiative mixing of non-thermal DM with light neutrinos. However, due to non-thermal nature of this DM candidate, the relevant Yukawa coupling $y^{\prime}_i$ in Eq.(\ref{yukawa2}) are very small, as we discuss in upcoming sections. For such tiny couplings, the mixing between non-thermal DM and light neutrinos will be too small to have any observable consequences like monochromatic lines in X-ray or Gamma-ray spectrum. We leave such exploration of detection prospects for such non-thermal DM candidates to future studies.

\begin{table}[htb]
\centering
\begin{tabular}{|c|c|c|}
\hline
Parameters & Normal Hierarchy (NH) & Inverted Hierarchy (IH) \\
\hline
$ \frac{\Delta m_{21}^2}{10^{-5} \text{eV}^2}$ & $6.79-8.01$ & $6.79-8.01 $ \\
$ \frac{|\Delta m_{31}^2|}{10^{-3} \text{eV}^2}$ & $2.427-2.625$ & $2.412-2.611 $ \\
$ \sin^2\theta_{12} $ &  $0.275-0.350 $ & $0.275-0.350 $ \\
$ \sin^2\theta_{23} $ & $0.418-0.627$ &  $0.423-0.629 $ \\
$\sin^2\theta_{13} $ & $0.02045-0.02439$ & $0.02068-0.02463 $ \\
$ \delta (^\circ) $ & $125-392$ & $196-360$ \\
\hline
\end{tabular}
\caption{Global fit $3\sigma$ values of neutrino oscillation parameters \cite{Esteban:2018azc}.}
\label{tabglobalfit}
\end{table}

\section{Analysis of co-genesis}
\label{sec3}
In order to do the entire analysis we need to solve the coupled differential equations of thermal DM (which is thermally produced and denoted by DM hereafter), non-thermal DM (denoted by $\chi$), lepton asymmetry as well as the source of $\chi$ (and partial source of lepton asymmetry) which is the lightest heavy right handed neutrino i.e. $N$ which provides a non-thermal origin of $\chi$. The coupled Boltzmann equations for DM and $N$ are given as

For this scenario the Boltzmann equations 
for the $Z_2$ odd particles take the following form:
\begin{align}
    \frac{dY_{N_k}}{dz} &= -\frac{s}{zH(z)}\left[(Y_{N_k}-Y^{\rm eq}_{N_k})\langle \Gamma_{N_k \rightarrow L_\alpha \eta}\rangle + (Y_{N_k}-Y^{\rm eq}_{N_k})\langle \Gamma_{N_k \rightarrow \phi \chi}\rangle \right. \nonumber \\
    &+ \left. (Y_{N_k}Y_{\eta}-Y^{\rm eq}_{N_k}Y^{\rm eq}_{\eta})\langle \sigma v \rangle_{\eta N_k\rightarrow L{\rm SM}} + \sum_{l=1}^3\left[(Y_{N_k}Y_{N_l}-Y^{\rm eq}_{N_k}Y^{\rm eq}_{N_l}) s\langle \sigma v \rangle_{N_l N_k\rightarrow {\rm SM SM}} \right. \right. \nonumber \\
    &+ \left. \left. s\langle\sigma v\rangle_{\rm N_kN_l\rightarrow \chi \chi} Y_{N_k}Y_{N_l} - (Y_{N_k}Y_{N_l} - Y^2_{\eta}\gamma^{N_k}_{\eta}\gamma^{N_l}_{\eta})s\langle \sigma v\rangle_{\rm N N \rightarrow \eta \eta}\right]\right], \nonumber \\
    \frac{dY_{\eta}}{dz} &= \frac{s}{zH(z)}\left[ 
    (Y_{N_k}-Y^{\rm eq}_{N_k})\langle \Gamma_{N_i \rightarrow L_\alpha \eta}\rangle - (Y_{\eta}-Y^{\rm eq}_{\eta})\langle \Gamma_{\eta \rightarrow L_\alpha \phi \chi}\rangle
     - 2(Y^2_{\eta}-(Y^{\rm eq}_{\eta})^2)\langle \sigma v\rangle_{\eta \eta \rightarrow {\rm SM SM}}\right. \nonumber \\
    &- \left. \sum^3_{m=1}(Y_{N_m}Y_{\eta}-Y^{\rm eq}_{N_m}Y^{\rm eq}_{\eta})\langle \sigma v \rangle_{\eta N_m\rightarrow L{\rm SM}} - (Y_{\phi}Y_{\eta}-Y^{\rm eq}_{\phi}Y^{\rm eq}_{\eta})\langle \sigma v \rangle_{\eta \phi\rightarrow {\rm SM}{\rm SM}} \right. \nonumber \\
    &+ \left. (Y_{N_k}Y_{N_l} -Y^2_{\eta}\gamma^{N_k}_{\eta}\gamma^{N_l}_{\eta})s\langle \sigma v\rangle_{\rm N N \rightarrow \eta \eta} \right]. \nonumber \\
    \frac{dY_{\phi}}{dz} &= \frac{s}{zH(z)}\left[(Y_{N_k}-Y^{\rm eq}_{N_k})\langle \Gamma_{N_k \rightarrow \phi \chi}\rangle + (Y_{\eta}-Y^{\rm eq}_{\eta})\langle \Gamma_{\eta \rightarrow L_\alpha \phi \chi}\rangle -2(Y^2_{\phi}-(Y^{\rm eq}_{\phi})^2)\langle \sigma v\rangle_{\phi \phi \rightarrow {\rm SM SM}} 
    \right. \nonumber \\
    &- \left.(Y_{\phi}Y_{\eta}-Y^{\rm eq}_{\phi}Y^{\rm eq}_{\eta})\langle \sigma v \rangle_{\eta \phi\rightarrow {\rm SM}{\rm SM}} \right].\label{eq:BE1}
    \end{align} 
where 
\begin{align}    
    Y^{\rm eq}_i &= n^{\rm eq}_i/s,~~s=g_*\frac{2\pi^2}{45}T^3,
    ~~H\sqrt{\frac{4\pi^3}{45}g_*}\frac{T^2}{M_{pl}}, \nonumber \\
    \langle \Gamma \rangle &= \Gamma \frac{K_1(z)}{K_2(z)},~~~~\gamma^i_j=\frac{n^{eq}_i}{n^{eq}_j},~~z=\frac{M_N}{T} \nonumber \\
      \langle \sigma_{ij\rightarrow kl} v \rangle &= \frac{x_f}{8m^2_im^2_jK_2((M_i/M_{N})x_f)K_2((M_j/M_{N})x_f)} \nonumber \\
    &\times\int_{s_{int}}^\infty \sigma_{ij\rightarrow kl}(s-2(M^2_i + M^2_j))\sqrt{s}K_1(\sqrt{s}x_f/M_{N})
\end{align}
and  $z=\frac{M_{\rm DM}}{T}$, $M_{\rm PL}$ is the Planck mass, $Y=n/s$ 
denotes  the ratio of number density to entropy density, $s_{\rm int} = \text{Max}\{(M_i+M_j)^2,(M_k+M_l)^2\}$, $M_{\rm Pl}$ is the Planck Mass,
$H$ is the Hubble rate of expansion, $n^{\rm eq}_i$'s are the equilibrium number density of 
$i^{th}$ species and $K_i$'s are the modified Bessel functions of order $i$. 
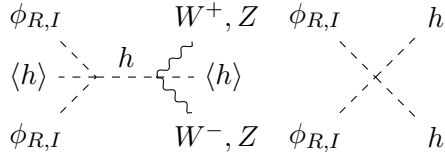
\begin{figure}
    \centering
    \begin{tabular}{lr}
        \begin{tikzpicture}[/tikzfeynman/small]
        \begin{feynman}
        \vertex (i){$\phi_{R,I}$};
        \vertex [below = 1.6cm of i] (j){$\phi_{R,I}$};
        \vertex [below = 0.8cm of i] (k);
        \vertex [right = 0.8cm of k] (v1);
        \vertex [left = 0.5cm of v1] (v1a){$\langle h\rangle$};
        \vertex [right = 0.8cm of v1] (v2);
        \vertex [right = 0.5cm of v2] (v2a){$\langle h\rangle$};
        \vertex [right = 2.4cm of i] (m){$W^+,Z$};
        \vertex [below = 1.6cm of m] (o){$W^-,Z$};
\diagram*[small]{(i) -- [scalar](v1),(j) -- [scalar](v1),(v1) -- [scalar,edge label=$h$](v2),(v2) -- [boson](m),(v2) -- [boson](o),(v1) -- [scalar](v1a),(v2) -- [scalar](v2a)};
\end{feynman}
\end{tikzpicture} &  
\begin{tikzpicture}[/tikzfeynman/small]
        \begin{feynman}
        \vertex (i){$\phi_{R,I}$};
        \vertex [below = 1.6cm of i] (j){$\phi_{R,I}$};
        \vertex [below = 0.8cm of i] (k);
        \vertex [right = 0.8cm of k] (v1);
        \vertex [right = 1.6cm of i] (m){$h$};
        \vertex [below = 1.6cm of m] (o){$h$};
\diagram*[small]{(i) -- [scalar](v1),(j) -- [scalar](v1),(v1) -- [scalar](m),(v1) -- [scalar](o)};
\end{feynman}
\end{tikzpicture}
    \end{tabular}
    \caption{Diagrams for the major annihilation channel responsible for the relic abundance of cold dark matter.}
    \label{fig:DMann}
\end{figure}
It is worthwhile to notice that the main annihilation proceeses leading to sufficient relic density for cold dark matter we consider are presented in FIG. \ref{fig:DMann}.

Now, in our model we have the possibility of generating the leptonic asymmetry 
through the co-annihilation channel of right handed $N_i$ and $\eta$ \cite{Borah:2018uci}. This is shown in FIG. \eqref{fig:coasym} where the doublet $\eta$ co-annihilates with $N_i$ into $\nu,X$ or $e^-,X$ where $X \equiv h,\gamma,W^{\pm},Z $. The $CP$-violation
comes from the interference with the loop diagram of the vertex. In addition to that, there can be additional contribution to lepton asymmetry from 
the decay of  the lightest right handed neutrino $N(\equiv N_1)$ as well, similar to vanilla leptogenesis.
The vanilla leptogenesis scenario with hierarchical $N_i$ masses~\cite{Buchmuller:2004nz} is
viable for the mass of $N_1$
satisfying $M_{1}^{\rm min}\gtrsim 10^9$ GeV~\cite{Davidson:2002qv, Buchmuller:2002rq}.\footnote{Including flavor and thermal effects could, in principle, lower this bound to about $10^6$ GeV~\cite{Moffat:2018wke}.
Addition of real scalar singlet can further reduce it to 500 GeV \cite{Alanne:2018brf}.}
 A similar lower bound can be derived in the scotogenic model with only two $\mathbb{Z}_2$ odd SM-singlet fermions in the strong washout regime. However, with three such SM-singlet fermions, the bound can be lowered to about 10 TeV~\cite{Hugle:2018qbw, Borah:2018rca}, even without resorting to a resonant enhancement of the CP-asymmetry~\cite{Pilaftsis:2003gt, Dev:2017wwc}. We do not show the details of vanilla leptogenesis here, but include that contribution into account for the final lepton asymmetry. 
Instead, we highlight the other feature 
from WIMP DM annihilation, as it connects the source of baryon asymmetry to the dark matter sector. 
For the details of vanilla leptogenesis, we refer to the above references where vanilla leptogenesis 
was studied in the context of type I seesaw as well as minimal scotogenic model. It should be noted that typically, the lowest scale of lepton number violation is more effective in creating lepton asymmetry. This scale, in our case is the scale of WIMP DM freeze-out, which lies below the right handed neutrino masses. However, if right handed neutrino masses are not very heavy compared to WIMP DM mass, then both of them can have some sizeable contribution to the origin of lepton asymmetry. We will show more details of our hybrid source of leptogenesis in a companion paper. 

The Boltzmann equations responsible for the  DM density is given by eq.(\ref{eq:BE1}) and leptonic asymmetry is given as follows:

\begin{align}
    \frac{dY_{\Delta L}}{dz} &=  \frac{s}{zH(z)}\left[\sum_i\left( \epsilon_{N_i}(Y_{N_i} - Y^{\rm eq}_{N_i})\langle \Gamma_{N_i \rightarrow L_\alpha \eta}\rangle - Y_{\Delta L}r_{N_i}\langle\Gamma_{N_i \rightarrow L_\alpha \eta}\rangle \right.\right. \nonumber \\
     &+ \left. \left. \epsilon_{N_i \eta} \langle \sigma v\rangle_{\eta N_i\rightarrow L {\rm SM}}\left(Y_{\eta}Y_{N_i} - Y^{\rm eq}_{\eta}Y^{\rm eq}_{N_i}\right) - \frac{1}{2}Y_{\Delta L}Y^{\rm eq}_lr_{N_i}r_\eta \langle \sigma v\rangle_{\eta N_i \rightarrow {\rm SM} \overline{L}} \right)\right.\nonumber \\
     &- \left. Y_{\Delta L}Y^{\rm eq}_lr^2_\eta\langle  \sigma v\rangle_{\eta \eta \rightarrow LL}  - Y_{\Delta L}Y^{\rm eq}_\eta\langle \sigma v \rangle^{wo}_{\eta L \rightarrow \eta \overline{L}} \right], \label{eq:asym} \\
     H &= \sqrt{\frac{4\pi^3 g_*}{45}}\frac{M^2_{\chi}}{M_{\rm PL}}, \quad s = g_* \frac{2\pi^2}{45}\left(\frac{M_{\chi}}{z}\right)^3, \nonumber \\
     r_j &= \frac{Y^{\rm eq}_j}{Y^{\rm eq}_l}, \quad  \quad
     \langle \Gamma_{j\rightarrow X} \rangle = \frac{K_1(M_j/T)}{K_2(M_j/T)}\Gamma_{j\rightarrow X}, \nonumber
\end{align}
And the CP asymmetry which is arising from the interference between tree and 1-loop diagrams in Fig. \ref{fig:coasym} can be estimated as

\begin{align}
\epsilon_{N_i \eta} &= \frac{1}{4\pi(yy^\dagger)_{ii}} \sum_{j}\Im[(yy^\dagger)^2_{ij}]\widetilde{\epsilon}_{ij}, \label{eq:asymB} \\
 \widetilde{\epsilon}_{ij}   &= \frac{\sqrt{r_j}}{6 r_i
   \left(-r_i^{3/2}+r_i (r_j-2)+\sqrt{r_i} r_j+1\right)^2(\sqrt{r_i}-3)} \left(r_i^{7/2} (3 r_j+1) +\sqrt{r_i} (3 r_j+5)+1 \right.\nonumber \\
   &- \left.3 r_i^{5/2} \left(r_j \left(D+(r_j-3) r_j+4\right)  -  3 D -2\right)-3 r_i^{3/2}
   \left(2 \left(D+3\right) + r_j \left(r_j \left(D+r_j+1\right)-D-4\right)\right) \right. \nonumber \\
   &- \left. r_i^4+f^3 \left(3
   D+3 r_j^2+11\right) - 3 r_i^2 \left(r_j \left(D+2 (r_j-1) r_j+2\right) - D +6\right)+r_i \left(1-3
   r_j \left(D+r_j-4\right)\right) \right) \nonumber \\
   &+ \frac{\sqrt{r_j}}{4 r_i}\left(\sqrt{r_i}-1+\frac{\sqrt{r_j}}{(1+\sqrt{r_i})^2}(\sqrt{r_i}-1+r_j)\left(\log\left(\frac{1+\sqrt{r_i}r_j}{r_i(1+\sqrt{r_i})}\right)-\log\left(\frac{1+r_i+r^{3/2}_i+\sqrt{r_i}r_j}{r_i(1+\sqrt{r_i})} \right)\right.\right. \nonumber \\
   &+ \left.\left. \log\left(1+\frac{1+\sqrt{r_i}}{\sqrt{r_i}(\sqrt{r_i}-1+r_i+r_j)}\right)\right)\right)\label{eq:coanneps} \\
   D &= \sqrt{(r_i-r_j) \left(r_i+4 \sqrt{r_i}-r_j+4\right)} \qquad
  r_l = \frac{M^2_{N_l}}{m^2_\eta}. \nonumber 
\end{align}

    It should be noted that in the above expression always ($1\leq r_j\leq r_i$) where $j$ stands for $N_j$ inside the loop while $i$ stands for $N_i$ as one of the initial state particles, shown in Fig. \ref{fig:coasym}. This is simply to realise the "on-shell" -ness of the loop particles in order to generate the required CP asymmetry. In the above Boltzmann equation we see that along with the process which produces the asymmetry i.e $\langle \sigma v \rangle_{N_i \eta \rightarrow X L}$ and $\langle \Gamma_{N_i \rightarrow L \eta} \rangle$ we have washout terms coming from three kinds of processes: 1)the process $\langle \sigma v \rangle_{\eta \eta \rightarrow L L}$  poses as one of the wash out along with 2) $\langle \sigma v \rangle_{\eta L \rightarrow \eta \overline{L}}$. Now, according to Cui {\it et. al} \cite{Cui:2011ab} if we need to achieve 
asymmetry through dark matter annihilation then the {\it wash-out} processes $\langle \sigma v \rangle_{N_i \eta \rightarrow X L}$ should {\it freeze-out} before the WIMP {\it freeze-out}. In order to do that one has to keep the following ratio below unity
\begin{align}
    \frac{\Gamma_{\rm wash-out}(x)}{\Gamma_{\rm WIMP}(x)}\sim \frac{\langle \sigma_{wash-out} v \rangle \prod_{i} Y^{eq}_i(x)}{4\langle \sigma_{ann} v\rangle Y^{eq}_X(x) Y_\gamma}.
\end{align}
So, in our case $\langle \sigma_{ann} v\rangle$ is similar to that of the standard Inert Doublet Model WIMP annihilation channel ($\eta \eta \rightarrow W^+ W^-$) which is naturally stronger than the $\langle \sigma_{wash-out} v \rangle$ which in our case is for ($N_i \eta \rightarrow X L$). Further details of the asymmetry generated through such t channel annihilations of dark matter are shown in \cite{Borah:2018uci}.

Non-thermal DM can be produced in a way similar to the FIMP scenario mentioned above. In such a case, the initial abundance is assumed to be zero or negligible and its interaction rate with the standard model particles or thermal bath is so feeble that thermal equilibrium is never attained. In such a case, non-thermal can be produced by out of equilibrium decays or scattering from particles in the thermal bath while the former typically dominates if same type of couplings is involved in both the processes. Further details of this mechanism for keV scale sterile neutrinos can be found in \cite{Merle:2015oja, Shakya:2015xnx, Konig:2016dzg} as well as the review on keV sterile neutrino DM \cite{Adhikari:2016bei} \footnote{A thermally produced keV scale sterile neutrino typically overcloses the universe \cite{Nemevsek:2012cd, Bezrukov:2009th, Borah:2017hgt}. This requires late time entropy dilution mechanism due to the late decay of heavier right handed neutrinos \cite{Scherrer:1984fd} or some kind of non-standard cosmological phase \cite{Biswas:2018iny}.}. For a general review of FIMP DM paradigm, please see \cite{Bernal:2017kxu}, as mentioned earlier.

Using the FIMP prescription described in the above-mentioned works, we can write down the corresponding Boltzmann equation for $\chi$, the FIMP candidate as
\begin{align}
    \frac{dY_\chi}{dz} &= \frac{1}{zH}\left[\sum_{i}\left(Y_{N_i}\langle \Gamma_{\rm N_i\rightarrow \phi \chi} \rangle + \sum_{j}Y_{N_i}Y_{N_j}s\langle \sigma v \rangle_{\rm N_iN_j\rightarrow \chi \chi} \right) + Y_{\eta}\langle \Gamma_{\eta \rightarrow L_\alpha \phi \chi}\rangle \right. \nonumber \\
    &+ \left. Y_lY_{\eta}s\langle \sigma v \rangle_{\eta l \rightarrow \phi \chi}\right].
\end{align}
Here the first contribution on the right hand side is from the decay process $N\rightarrow \phi\chi$ while the second one is from annihilation $N N \rightarrow \chi \chi$. The fact that $\chi$ was never produced in equilibrium requires the Yukawa coupling governing the interaction among $N, \phi, \chi$ to be very small, as we mention below. Since the same Yukawa coupling appears twice in the annihilation process $N N \rightarrow \chi \chi$, the two body decay will dominate the production. Another dominant contribution can come from the $s$-channel annihilation process of $(\nu,l^{\pm}),(\eta^0,\eta^\pm)\rightarrow \phi \chi$ that appears in the third term on the right hand side of the above equation. The dominant production processes of $\chi$ in our work are shown in FIG.  \ref{fig:fimp}.
\begin{figure}
\centering
\includegraphics[width=0.65\textwidth]{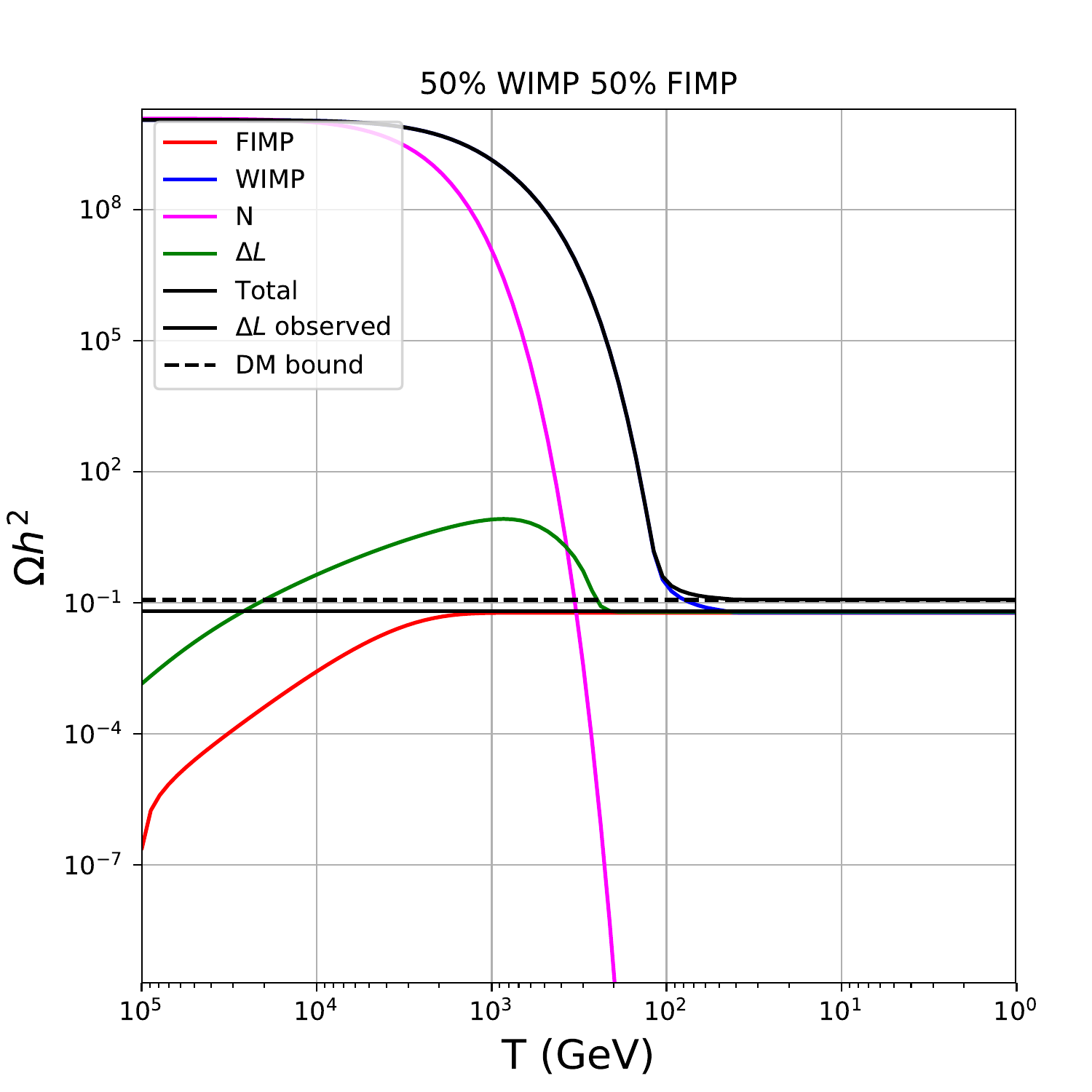} 
\caption{Co-genesis of equal abundance of WIMP (3 TeV mass) and FIMP (1 keV mass) DM candidates and baryon asymmetry.} 
\label{cogen1}
\end{figure}

\begin{figure}
\centering
\includegraphics[width=0.65\textwidth]{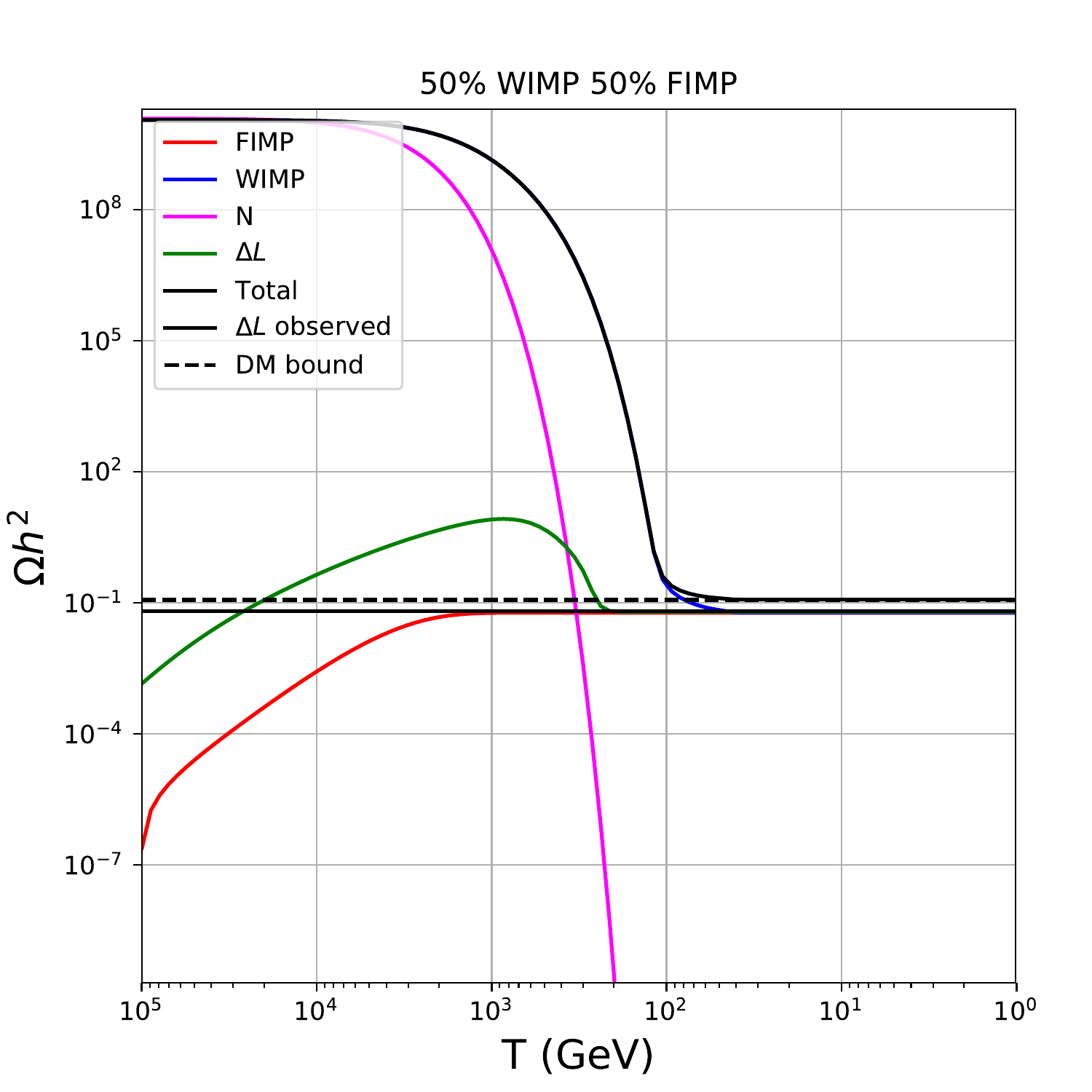} \\
\caption{Co-genesis of equal abundance of WIMP (3 TeV mass) and FIMP (1 MeV mass) DM candidates and baryon asymmetry.} 
\label{cogen2}
\end{figure}

\begin{figure}
\centering
\includegraphics[width=0.65\textwidth]{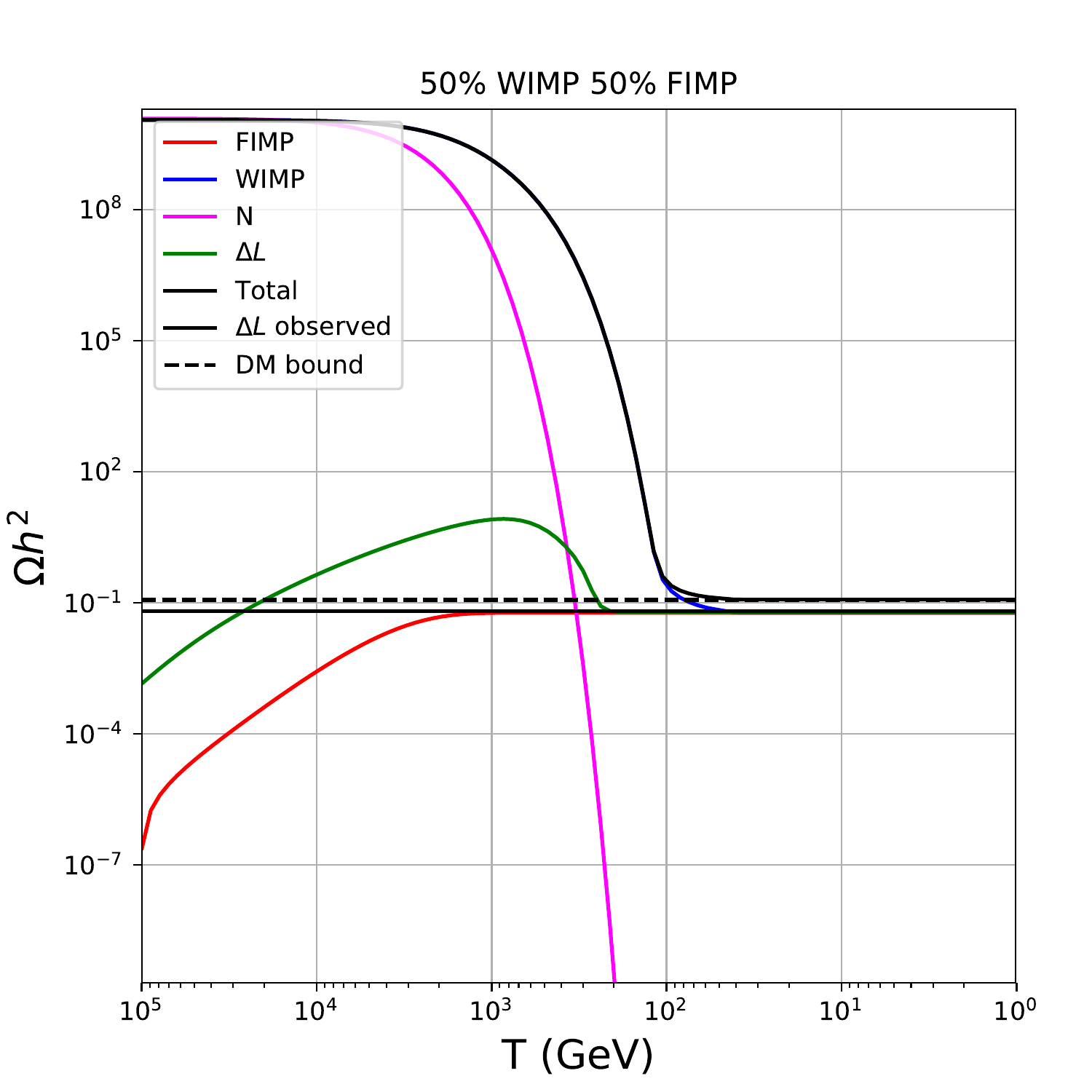} 
\caption{Co-genesis of equal abundance of WIMP (3 TeV mass) and FIMP (1 GeV mass) DM candidates and baryon asymmetry.} 
\label{cogen3}
\end{figure}
\section{Direct Detection and Lepton Flavour Violation}
\label{sec4}
Although the detection prospects of FIMP candidate are very limited, the WIMP can have very good direct detection signatures that can be probed at direct detection experiments like LUX \cite{Akerib:2016vxi}, PandaX-II \cite{panda17, Tan:2016zwf} and Xenon1T \cite{Aprile:2017iyp, Aprile:2018dbl}. Since the WIMP is a scalar, we can have Higgs mediated spin independent elastic scattering of DM off nucleons. This direct detection cross section can be estimated as \cite{Barbieri:2006dq}
\begin{equation}
 \sigma_{\text{SI}} = \frac{\lambda^2_L f^2}{4\pi}\frac{\mu^2 m^2_n}{m^4_h m^2_{\rm DM}}
\label{sigma_dd}
\end{equation}
where $\mu = m_n m_{\rm DM}/(m_n+m_{\rm DM})$ is the DM-nucleon reduced mass and $\lambda_L$ is the quartic coupling involved in DM-Higgs interaction. For WIMP, an admixture of scalar doublet and scalar singlet given by $\eta_1 = \cos{\theta_2} \eta_R + \sin{\theta_2} \phi$, the Higgs-DM coupling will be $\lambda_L = \cos{\theta_2} (\lambda_3+\lambda_4+\lambda_5) + \sin{\theta_2} \lambda_7/2$. A recent estimate of the Higgs-nucleon coupling $f$ gives $f = 0.32$ \cite{Giedt:2009mr} although the full range of allowed values is $f=0.26-0.63$ \cite{Mambrini:2011ik}. Since DM has a doublet component in it, there arises the possibility of tree level $Z$ boson mediated processes $\eta_R n \rightarrow \eta_I n$, $n$ being a nucleon. This process, if allowed, can give rise to a very large direct detection rate ruled out by experimental data. However, due to the inelastic nature of the process, one can forbid such scattering if $\delta = m_{\eta_I} - m_{\eta_R} > 100$ keV, typical kinetic energy of DM particle.

Another interesting observational prospect of our model is the area of charged lepton flavour violation. In the SM, there is no such process at tree level. However, at radiative level, such processes can occur in the SM. But they are suppressed by the smallness of neutrino masses, much beyond the current and near future experimental sensitivities. Therefore, any experimental observation of such processes is definitely a sign of BSM physics, like the one we are studying here. In the present model, this becomes inevitable due to the couplings of new $\mathbb{Z}_2$ odd particles to the SM lepton doublets. The same fields that take part in the one-loop generation of light neutrino mass shown in FIG. \ref{numassmixing} can also mediate charged lepton flavour violating processes like $\mu \rightarrow e \gamma, \mu \rightarrow 3e $ etc. For example, the neural scalars in the internal lines of loops in FIG. \ref{numassmixing} will be replaced by their charged counterparts (which emit a photon) whereas the external fermion legs can be replaced by $\mu, e$ respectively, giving the one-loop contribution to $\mu \rightarrow e \gamma $. Since the couplings and masses involved in this process are the same as the ones that generate light neutrino masses and play a role in DM relic abundance, we can no longer choose them arbitrarily. Lepton flavour violation in scotogenic model was studied by several authors including \cite{Vicente:2014wga, Toma:2013zsa}. 

Here we use the \texttt{SPheno 3.1} interface to check the constraints from cLFV data. We particularly focus on three such cLFV decays namely, $\mu \rightarrow e \gamma, \mu \rightarrow 3e$ and $\mu \rightarrow e$ (Ti) conversion that not only are strongly constrained by present experiments but also have tantalising future prospects \cite{Toma:2013zsa}. The present bounds are: ${\rm BR}(\mu \rightarrow e \gamma) < 4.2 \times 10^{-13}$ \cite{TheMEG:2016wtm},  ${\rm BR}(\mu \rightarrow 3e) < 1.0 \times 10^{-12}$ \cite{Bellgardt:1987du}, ${\rm CR} (\mu, \rm Ti \rightarrow e, \rm Ti) < 4.3 \times 10^{-12}$ \cite{Dohmen:1993mp}. 
It may be noted that the sensitivities of the first two processes will be improved by around one order of magnitude compared the present upper limit on branching ratios. On the other hand, the $\mu$ to $e$ conversion (Ti) sensitivity will be increased by six order of magnitudes \cite{Toma:2013zsa} making it a highly promising test of different new physics scenarios around the TeV corner.
\begin{figure}[!h]
\centering
\begin{tabular}{cc}
\includegraphics[width=0.50\textwidth]{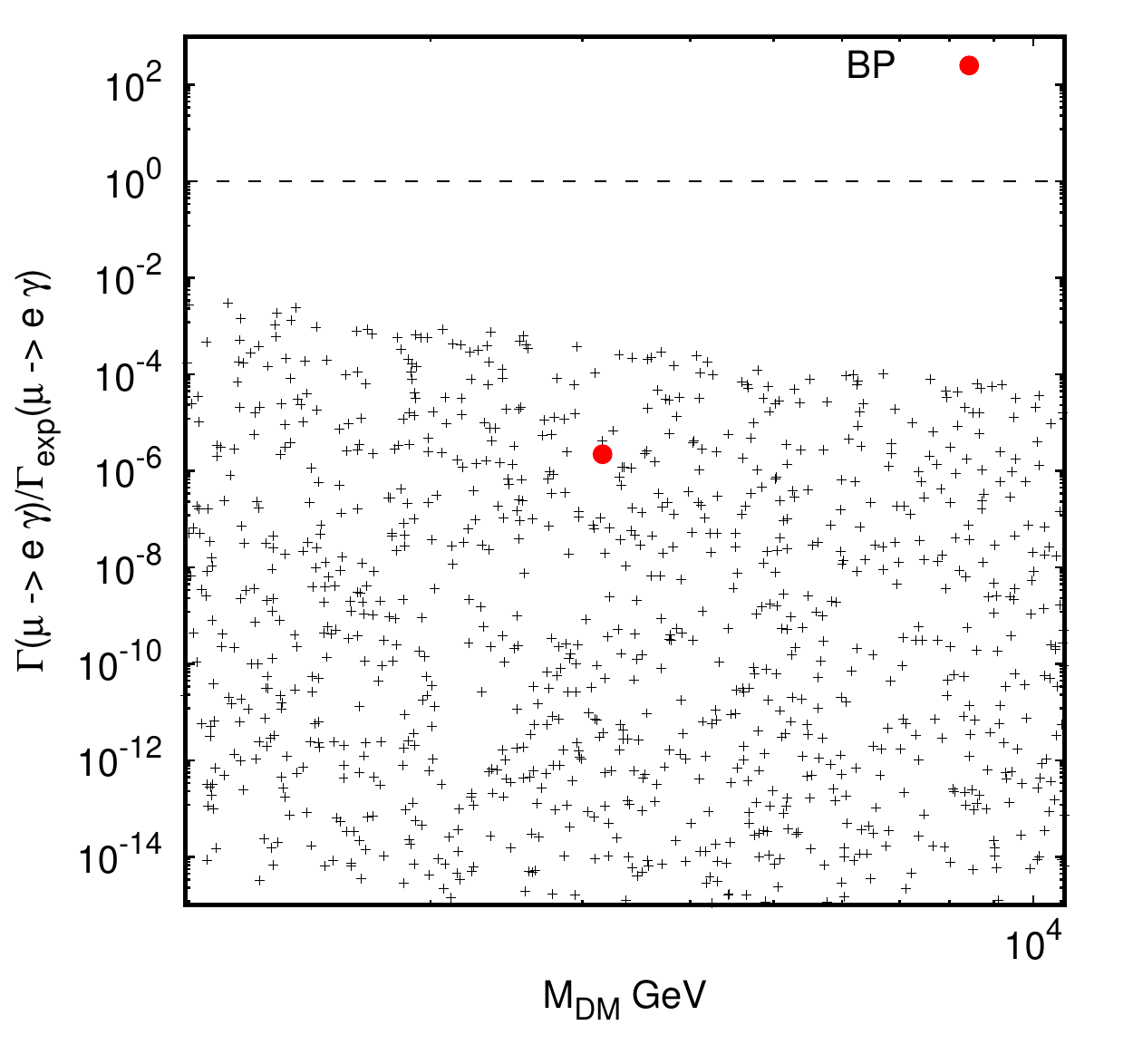} &
\includegraphics[width=0.50\textwidth]{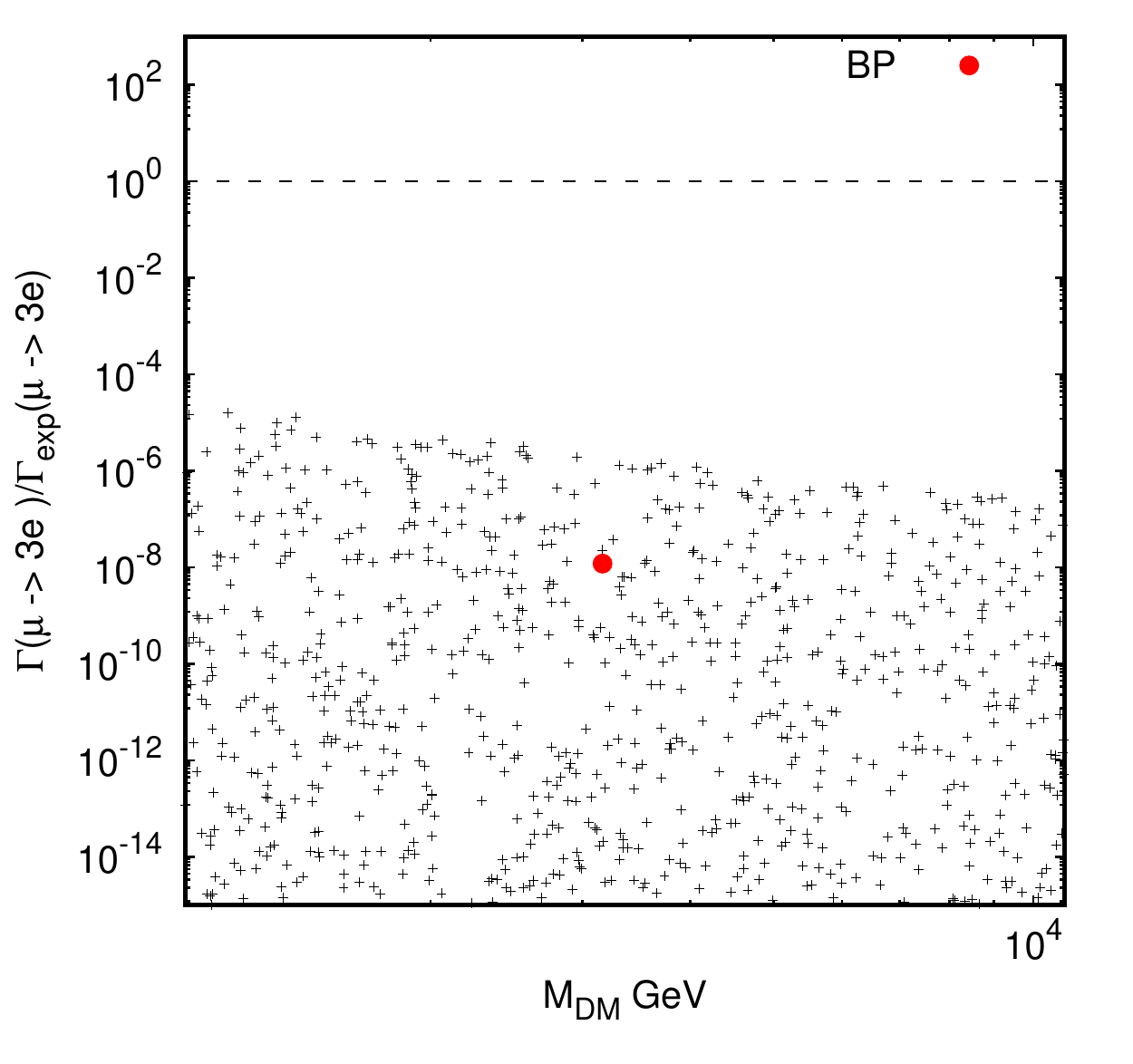} \\
\includegraphics[width=0.50\textwidth]{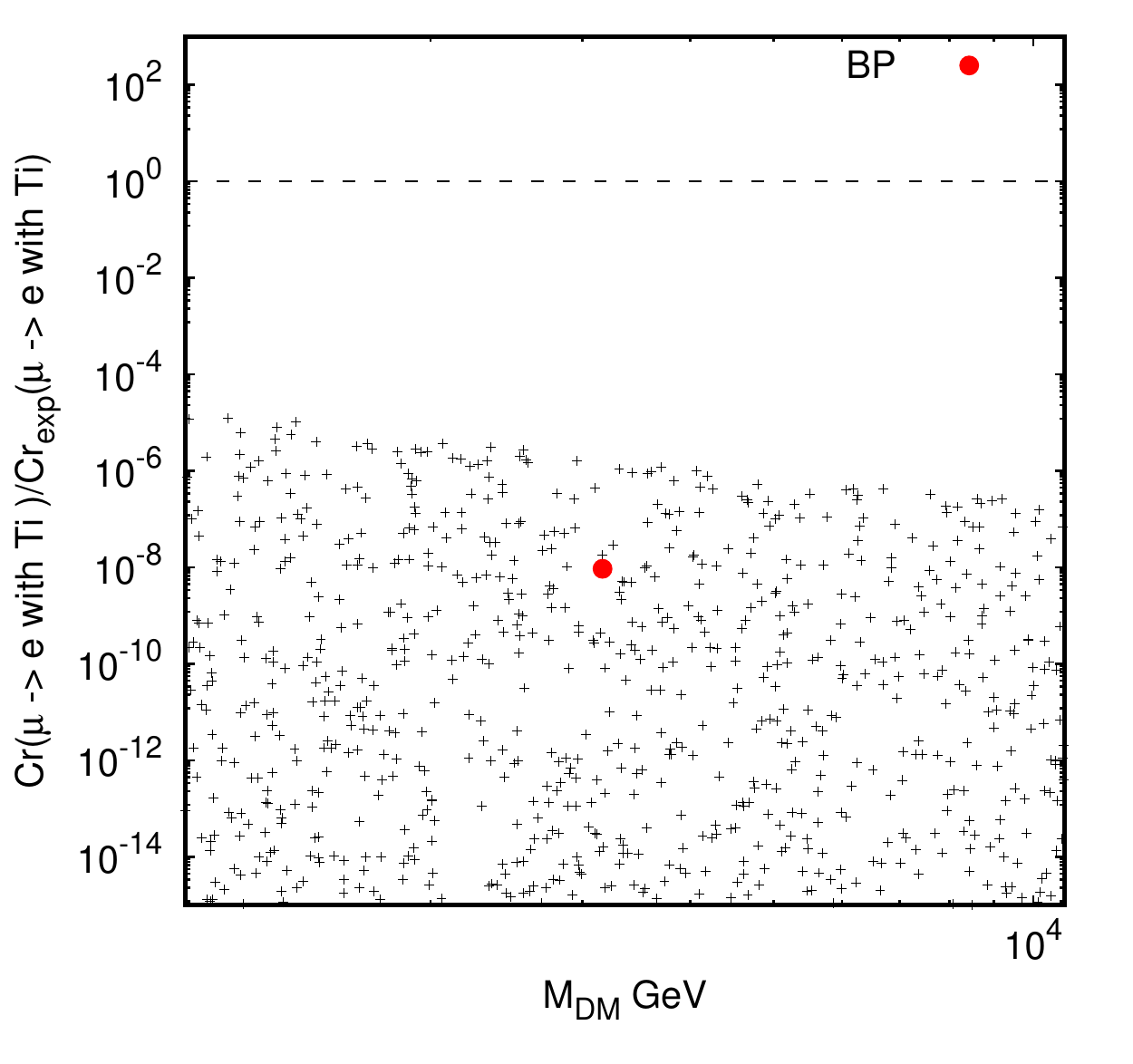} 
\end{tabular}
\caption{Scatter plots for three LFV processes obtained for the range of parameters mentioned in table \ref{tab:BPb}. The y axes correspond to
the ratio of the predicted decay rate to the experimental upper limit.
The black dots correspond to our benchmark point(BP) as given in table \ref{tab:BPa} and  satisfy all relevant bounds from WIMP-FIMP as well as correct lepton asymmetry. The horizontal dashed lines correspond to the experimental upper bound.}
\label{fig:scatt}
\end{figure}

\begin{figure}[!h]
\centering
\includegraphics[width=0.50\textwidth]{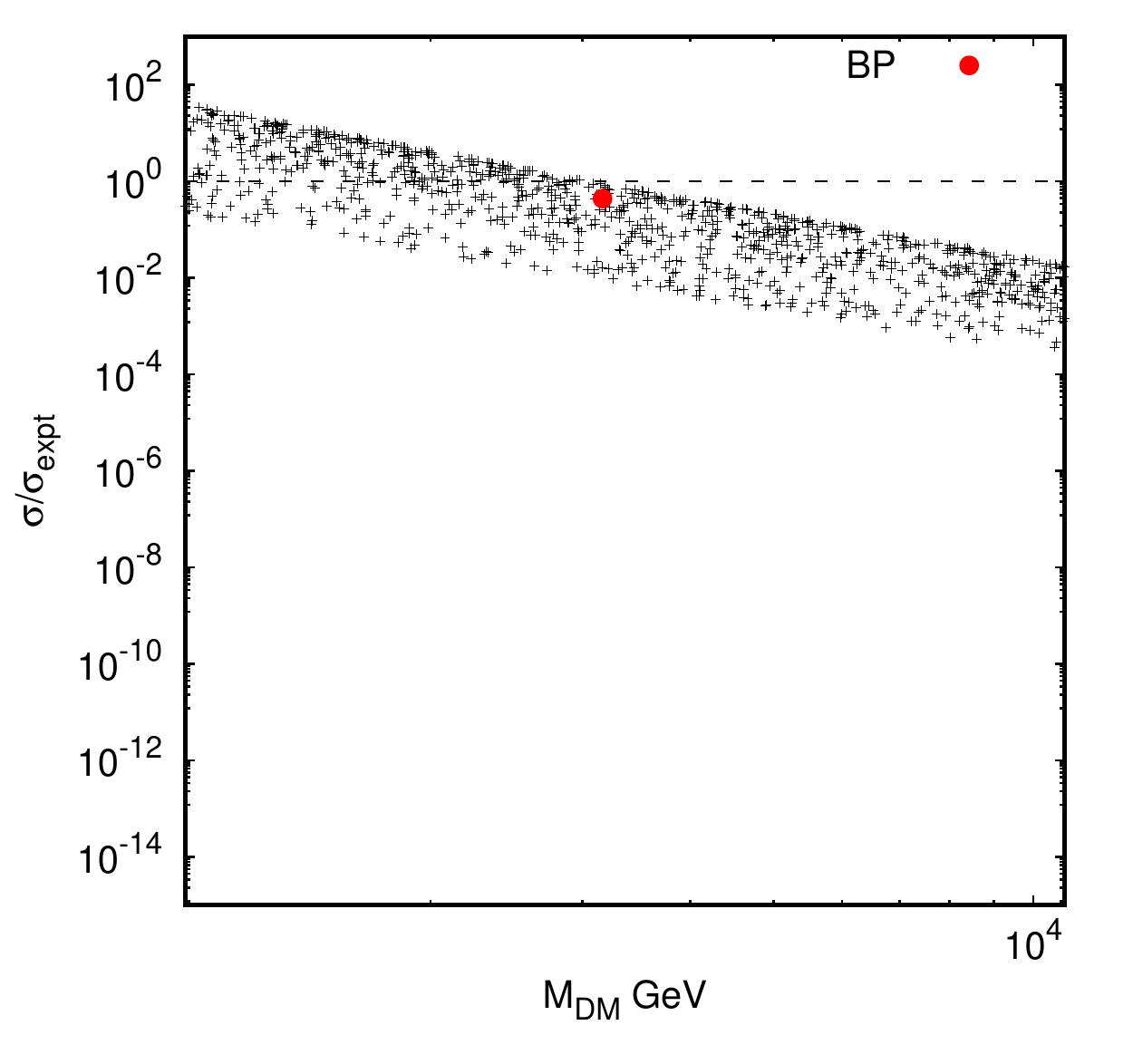}
\caption{The scatter plot for the ratio of spin independent direct detection rates for WIMP DM to the experimental bound for the range of parameters mentioned in table \ref{tab:BPb}. 
The black dot corresponds to our benchmark point (BP) corresponds to the parameters given in table \ref{tab:BPa} and satisfies all relevant bounds from WIMP-FIMP as well as correct lepton asymmetry. The horizontal dashed line corresponds to the experimental upper bound.}
\label{fig:dd}
\end{figure}

\section{Results and Discussion}
\label{sec5}
Since we have a large parameter space, we first choose the benchmark points in a way that gives rise to the desired phenomenology. Also, we choose three different masses of FIMP DM namely, 1 keV, 1 MeV and 1 GeV and choose the parameters in such a way that all these three cases correspond to $50\%$ contribution of FIMP to total DM abundance. The WIMP DM mass is kept fixed at 3 TeV in our analysis. We find that this relative contribution of FIMP along with the required lepton (and hence baryon) asymmetry can be generated by varying the yukawa coupling $y'_i$ as shown in table \ref{tab:BP}. In fact $y'_1$ solely decides the abundance of FIMP for a particular mass as this is the only parameter through which FIMP can couple to other particles in the model. As can be seen from this table, one requires very small $y'_1$ for the FIMP, as expected from the non-thermal scenario discussed earlier. One also requires relatively large quartic coupling $\lambda_7$ which decides the Higgs portal interactions of $\phi$ which is present as a small component in the WIMP DM eigenstate. The corresponding co-genesis results are shown in FIG. \ref{cogen1}, \ref{cogen2}, \ref{cogen3} for three different FIMP masses proportions respectively. As we can see from these plots, the WIMP as well as the lightest right handed neutrino are in equilibrium initially followed by WIMP freeze-out and right handed neutrino decay \footnote{Assumption of right handed neutrino to be in thermal equilibrium initially is justified in scotogenic model, as shown in \cite{Borah:2018rca}.}. Since the mass hierarchy is not very large, both the WIMP freeze-out and right handed neutrino abundance depletion (due to its decay) happens around the same epoch. Since WIMP freeze-out and right handed neutrino decay are related to the generation of lepton asymmetry as well as FIMP generation respectively, one can see the yield in $\Delta L$ and FIMP by the epochs of WIMP freeze-out and right handed neutrino decay. It can be seen from these plots that the required asymmetry along with WIMP-FIMP relative abundance can be achieved simultaneously leading to a successful co-genesis. In order to get the leptonic asymmetry we need the yukawa coupling $y_{ij}$
to be of $\mathcal{O}(1)$ which would be fulfilled if we take the $\lambda_5$ to be very less, to be in agreement with light neutrino masses discussed above. In doing so we would be compromising the mass difference between the $\eta_R$ and $\eta_I$. The decreasing of the mass difference opens up the inelastic channel $\eta_R,(n,p)\rightarrow \eta_I,(n,p)$ which is ruled out, as mentioned earlier. This is where the singlet scalar $\phi$ comes to rescue as it relaxes the tension among neutrino data, dark matter direct detection and generating correct lepton asymmetry This was also noted in a recent work \cite{Borah:2018uci}. The mixing between the doublet and singlet scalars through $\lambda_9$ helps in evading the Direct-Detection bound as is enters the effective $Z$-coupling to scalar WIMP. All these cases shown in FIG. \ref{cogen1}, \ref{cogen2}, \ref{cogen3} satisfy the final leptonic asymmetry (by the epoch of electroweak phase transition temperature $\sim 150-200$ GeV) required for the observed baryon asymmetry ($\eta_B = \Omega_B h^2 =0.0226$) via the sphaleron conversion factor $C_s = \frac{8N_f + 4N_H}{22N_f+13N_H}$ where $N_f=3, N_H=2$ are the number of fermion generations and Higgs doublets respectively.

It should be noted that the FIMP DM with mass in the keV scale can face constraints from structure formation data. As noted in \cite{Boyarsky:2008xj}, Lyman-$\alpha$ bounds restrict the keV fermion mass to be above 8 keV if it is non-resonantly produced (similar to our model) and contributes $100\%$ to the total DM abundance. However, for less than $60\%$ contribution to total DM, such strict mass bounds do not apply. Therefore, our benchmark value of 1 keV FIMP mass in one of the cases mentioned above remains safe from such bounds.

In table \ref{tab:BPa} we show the other parameters of the model for a chosen benchmark point (BP) giving $50\%-50\%$ WIMP-FIMP proportion along with successful leptogenesis. We will compare our subsequent results with respect to this BP that satisfies all our criteria. We will see that this BP remains sensitive to LFV as well as direct detection experiments.

\begin{table}[!ht]
\begin{tabular}{|c|c|}
\hline
FIMP mass & $y'_i$ \\ \hline
1 keV &  $5.4128\times 10^{-8}$   \\ \hline
1 MeV &  $1.711678\times 10^{-9}$   \\ \hline
1 GeV &  $5.4128\times 10^{-11}$   \\ \hline
\end{tabular}
\caption{Three different cases for FIMP mass and $y'_i$.}
\label{tab:BP}
\end{table}

\begin{table}[!ht]
\begin{tabular}{|c|c|}
\hline
Parameters & Values  \\ \hline
$\lambda_1$ & $0.17$   \\ \hline
$\lambda_5$ & $1.5\times 10^{-7}$   \\ \hline
$\lambda_3=\lambda_4=\lambda_9=\lambda^\prime_8$ & $0.1$    \\ \hline
$\lambda_7$ & $2.289$   \\ \hline
$\lambda^\prime_7=\lambda_8=\lambda^\prime_8=\lambda^\prime_9$ & $0.0$ \\ \hline
$\mu_\eta$ & $5.1$ TeV   \\ \hline
$\mu_\phi$ & $3$ TeV   \\ \hline
$m_\eta$ & $3.167$ TeV \\ \hline
$m_\phi$ & $5.298$ TeV \\ \hline
$M_N$ & $10.2$ TeV   \\ \hline
\end{tabular}
\caption{The benchmark point satisfying correct DM-leptogenesis requirements corresponding to $50\%-50\%$ relative proportion of WIMP-FIMP mentioned in table \ref{tab:BP}.}
\label{tab:BPa}
\end{table}

In FIG. \ref{fig:scatt} we have shown the scatter plot for LFV branching ratios by varying the key parameters affecting them, as shown in table \ref{tab:BPb}. In all these plots 
we have not taken any constraint from the relic, but the neutrinos mass constraints are being taken care of by the Casas-Ibarra parametrsation which in turn fixes the Yukawa's. The benchmark point that satisfies all relevant bounds from WIMP-FIMP as well as correct lepton asymmetry is also indicated as BP. For the same range of parameters we also show the WIMP direct detection rates in FIG. \ref{fig:dd} where the BP is also indicated. It is clear that our BP is very sensitive to the current experimental upper bounds on $\mu \rightarrow e \gamma$ as well as direct detection rates, keeping the detection prospects very much optimistic.

\begin{table}[!ht]
\begin{tabular}{|c|c|}
\hline
Parameter & variation \\ \hline
$\mu_\phi$ & 1 TeV - 10 TeV   \\ \hline
$\mu_\eta$ & 1.7 TeV - 17 TeV    \\ \hline
$M_N$ & 2.21 TeV - 22.1 TeV    \\ \hline
$\lambda_9$ & $10^{-3}$ - 1 \\ \hline
$\lambda_7$ & $10^{-1}$ - $4 \pi$ \\ \hline
\end{tabular}
\caption{Ranges of the parameters varied in order to get the scatter plots in FIG. \ref{fig:scatt}, \ref{fig:dd}.}
\label{tab:BPb}
\end{table}

\section{Conclusion}
\label{sec6}
We have studied the possibility of two-component dark matter with one thermal and one non-thermal components with the additional feature of creating the baryon asymmetry of the universe in a minimal extension of the standard model that also accommodates light neutrino masses radiatively with dark matter particles going inside loop. The model is a simple extension of the minimal scotogenic model which consists of the SM particles plus three right handed neutrinos, one additional scalar doublet in order to achieve the additional features, not present in the minimal model. The WIMP dark matter component is produced thermally in equilibrium followed by freeze-out while the non-thermal (or FIMP) component is produced from the out-of-equilibrium decay and scattering of particles in thermal bath. The WIMP annihilations also produce a non-zero lepton asymmetry in a way similar to WIMPy leptogenesis scenarios. The WIMP is an admixture of a scalar doublet's neutral component and a scalar singlet to satisfy the criteria of neutrino mass, dark matter relic, direct detection and leptogenesis simultaneously. Interestingly, the particles which assist in the production of FIMP also partially contribute to the origin of lepton asymmetry resulting in a hybrid setup. We outline such a hybrid co-genesis of multi-component DM, lepton asymmetry in this work for some benchmark scenarios leaving a more detailed analysis for an upcoming work. We also find that our benchmark point satisfying the required abundance of WIMP-FIMP and baryon asymmetry also remains sensitive to dark matter direct detection as well charged lepton flavour violation like $\mu \rightarrow e \gamma$.
\appendix
\acknowledgments
One of the authors, DB acknowledges the hospitality and facilities provided by School of Liberal Arts, Seoul-Tech, Korea where this work was initiated and to Harish-Chandra Research Institute Allahabad, India where part of this work was finalised. DB also acknowledges the support from IIT Guwahati start-up grant (reference number: xPHYSUGI-ITG01152xxDB001), Early Career Research Award from DST-SERB, Government of India (reference number: ECR/2017/001873) and Associateship Programme of IUCAA, Pune. SK and AD were supported by the National Research Foundation of Korea (NRF) grants (2009-0083526, 2017K1A3A7A09016430, 2017R1A2B4006338).

\bibliographystyle{apsrev}
\providecommand{\href}[2]{#2}\begingroup\raggedright
\endgroup
\end{document}